\documentclass[8pt]{article}
\usepackage{graphicx}
\usepackage{amssymb}
\def\beq{\begin{eqnarray}}
\def\eeq{\end{eqnarray}}

\usepackage{amsfonts}

\begin{document}


\title{Weakly bound states in heterogeneous waveguides: a calculation to fourth order}
\author{Paolo Amore \\
\small Facultad de Ciencias, CUICBAS, Universidad de Colima,\\
\small Bernal D\'{i}az del Castillo 340, Colima, Colima, Mexico \\
\small paolo.amore@gmail.com }

\maketitle

\begin{abstract}
We have extended a previous calculation of the energy of a weakly heterogeneous waveguide to fourth order in the density perturbation, 
deriving its general expression. For particular configurations where the second and third orders both vanish, we discover that the 
fourth order contribution lowers in general the energy of the state, below the threshold of the continuum. In these cases the waveguide 
possesses a localized state. We have applied our general formula to a solvable model with vanishing second and third orders reproducing 
the exact expression for the fourth order.
\end{abstract}

\section{Introduction}
\label{sec:Intro}
 
It is nowadays a well--known fact that bound states can appear in infinite waveguides  or tubes, 
in presence of an arbitrarily weak bending or of a local, small, enlargement of its section. 
This behavior has been proved for general configurations in Refs.~\cite{Exner89,Goldstone92} and it has
been investigated for several  specific geometrical configurations. It is impossible to refer to all 
the different works, but we would like to mention the case of the infinite symmetric cross studied by 
Schult and collaborators in Ref.~\cite{Schult89}. Although Ref.~\cite{Schult89} is focussed on 
the study of the quantum mechanical bound states of the symmetric cross, the problem is relevant in many 
areas of Physics, such as Acoustics, Electromagnetism and Fluid dynamics (in this respect, it is important 
to cite the work by Ursell Ref.\cite{Ursell51,Ursell52} who studied the emergence of trapped modes in a 
semi-infinite canal of fixed width terminating in a sloping beach). 
It is also important to mention that the appearance of bound states in waveguides and, more in general, in 
open geometries, must affect the transport properties of the systems, modifying the transmission and 
reflection coefficients (see for instance Ref.~\cite{Qu04}). 

From a mathematical point of view, one needs to solve the Helmholtz equation on an open, infinite domain, 
with Dirichlet  boundary conditions  at the border. In particular, Bulla and collaborators have considered 
in Ref.~\cite{Simon97} the problem of an infinite homogeneous waveguide on the region 
\begin{eqnarray}
\Omega_\lambda = \left\{ (x,y) \in {\mathbb{R}}^2 | 0 < y < \lambda f(x)\right\}
\end{eqnarray}
obeying Dirichlet boundary conditions at the border, assuming that $f$ is a $C^\infty ({\mathbb{R}})$ function 
of compact support with $f \geq 0$. In their calculation $\lambda>0$ is a parameter which controls the deformation 
of the border (particularly the case $\lambda=0$ reduces to a straight waveguide, with a purely continuum spectrum).
These authors were able to show that, if $\int_{-\infty}^{\infty} f(x) dx >0$, there is at least one eigenvalue
falling below the continuum threshold. They also obtained the exact expression for the energy of the fundamental
mode, to second order in the parameter controlling the deformation. Soon after, Exner and Vugalter \cite{Exner97}
studied this problem, when the deformation of the border averages out, i.e. when $\int_{-\infty}^{\infty} f(x) dx = 0$.
Interestingly they found out that under certain conditions it is still possible to have a bound state and that the 
energy gap scales as the fourth power in $\lambda$.

Recently, the present author and collaborators have studied in  Ref.~\cite{Amore16} a different
but related problem: the case of a infinite straight waveguide containing a small inhomogeneity 
centered at an internal point (assuming Dirichlet boundary conditions at the border). 
In that case, it was proved that, when the heterogeneity corresponds to a locally denser
region, the eigenfunction of the ground state becomes localized around the heterogeneity and the
corresponding energy falls below the continuum threshold. The calculation of Ref.~\cite{Amore16}
was carried out using perturbation theory up to third order, using an approach originally 
proposed by Gat and Rosenstein in Ref.~\cite{Gat93} for a different problem. As a matter of fact, 
the implementation of the perturbative scheme must be done with care, since the naive identification
of the unperturbed operator with the negative Laplacian would lead to the appearance of divergent 
contributions in the coefficients of the  perturbative series for the energy of the ground state. 
The emergence of these (infrared) divergences can be easily understood since the spectrum 
of $(-\Delta)$ on an infinite strip is continuous and therefore the denominators of the coefficients 
in the Rayleigh-Schr\"odinger expansion may become arbitrarily small. To avoid this problem in 
Ref.~\cite{Amore16} a suitable unperturbed operator was used, following the approach of
Gat and Rosenstein: the spectrum of this operator contains now a localized state and the continuum,
with the energy of the localized state falling below the continuum threshold (the separation
between the two depends on a parameter $\beta$ in the unperturbed operator which will be eventually 
set to zero). In this way one is able to carry out the usual perturbative expansion,
obtaining explicit expressions which are finite when $\beta \rightarrow 0^+$. 

In this paper we have extended the calculation of Ref.~\cite{Amore16}, obtaining the 
exact general expression for the energy correction to fourth order in the density perturbation. 
The greater technical difficulty of the present calculation derives both because from the larger number of
terms and both from their different nature. Working in our perturbation scheme we find
that all the infrared divergent terms (i.e. terms which diverge as $\beta \rightarrow 0^+$) potentially 
contained in $E_0^{(4)}$ correctly  cancel out, as expected.
Moreover, for the case where the second and third order corrections both vanish, we find that there is 
a non--vanishing fourth order correction to the energy of the fundamental mode, which lowers
the energy below the continuum threshold. Since the problem of Bulla et al. \cite{Simon97} may be converted
to the problem of an infinite heterogeneous waveguide, using a suitable conformal map, our results 
also provide an alternative approach to the problems studied in Refs.\cite{Simon97} and \cite{Exner97}. 
Additionally, our formulas apply as well to the case of infinite heterogeneous and deformed waveguides 
(in this case the "density" in  our formulas would involve both the physical density of the waveguide and 
the "conformal density" obtained from the mapping), thus allowing to treat more general problems.

The paper is organized as follows: in Section \ref{sec:PT} we discuss the perturbation theory, and present
the general formulas for the energy to fourth order; in Section \ref{sec:model} we consider a solvable 
model, reproducing the exact results to fourth order; in Section \ref{sec:Concl} we present our conclusions.
The Appendices \ref{appa} and \ref{appb} contain technical details of the calculation.

\section{Perturbation theory}
\label{sec:PT}

In a recent paper we have obtained the explicit expression for the energy of the fundamental mode of 
an infinite, weakly heterogeneous two dimensional waveguide, up to third order in the density perturbation. It is assumed
that the inhomogeneity is small and localized at some internal point of the waveguide. Under these assumptions
it is proved that, when the perturbation corresponds to a locally denser material, a bound state, localized
at the inhomogeneity appears. 

Mathematically, we are considering the Helmholtz equation
\begin{eqnarray}
\left( - \Delta \right) \Psi_n\left(\mathbf{x} \right) = E_n \Sigma\left(\mathbf{x} \right) \Psi_n\left(\mathbf{x} \right)
\label{eq_helmholtz_a}
\end{eqnarray}
where $|x| < \infty$ and $|y| \leq b/2$. The solutions obey Dirichlet boundary conditions at the border
\begin{eqnarray}
\Psi_n(x, \pm b/2) = 0
\end{eqnarray}
and $\Sigma(x,y) > 0$ for $|x| < \infty$ and $|y| \leq b/2$.

Expressing the density as $\Sigma\left(\mathbf{x} \right) = 1 + \sigma\left(\mathbf{x} \right)$,
where $\lim_{|x| \rightarrow \infty} \sigma\left(\mathbf{x} \right) = 0$, and assuming that $|\sigma(x)| \ll 1$ for $x \in (-\infty,\infty)$,
we can perform a perturbative expansion in the density perturbation.

The general formulas for the perturbative corrections to the energy of the fundamental mode 
up to third order have been derived in Refs.~\cite{Amore16} and \cite{Amore10b}  and read
\begin{eqnarray}
E_0^{(1)} &=& -\langle\sigma \rangle \epsilon _0 \\
E_0^{(2)} &=& \langle\sigma \rangle^2 \epsilon _0-\langle\sigma \Omega \sigma \rangle \epsilon _0^2 \\
E_0^{(3)} &=& -\epsilon _0 \langle\sigma \rangle^3 +3 \langle\sigma \rangle \langle\sigma \Omega \sigma \rangle \epsilon_0^2 + \epsilon _0^3 (\langle\sigma \rangle \langle\sigma \Omega \Omega \sigma \rangle-\langle\sigma \Omega \sigma
   \Omega \sigma \rangle) 
\end{eqnarray}   
where
\begin{eqnarray}
\hat{\Omega} \equiv \sum_{n} \frac{|n\rangle \langle n|}{\epsilon_n -\epsilon_0}
\end{eqnarray}
and $\epsilon_n$ and $|n\rangle$ are the eigenvalues and eigenstates of the unperturbed operator\footnote{In the following we
will adopt the notation $\langle \hat{A} \rangle$ to indicate the expectation value of the operator $\hat{A}$ in the
ground state of $\hat{H}_0$.}.

As we have discussed in Ref.~\cite{Amore16}, the identification of the unperturbed operator must be done with care, for the case
of an infinite waveguide: as a matter of fact, the obvious candidate, corresponding to an infinite, straight and homogeneous waveguide 
cannot be used, since its spectrum is continuous and the fundamental mode can thus be excited to states which are arbitrarily 
close in energy. In this case, the perturbative formulas would contain infrared divergences, which would completely spoil 
the calculation. In a different context Gat and Rosenstein \cite{Gat93} have devised a perturbation scheme that allows
to avoid these infrared divergences: in our case this process amounts to use as unperturbed operator
\begin{equation}
\hat{H}_0 = -\Delta - 2 \beta \delta(x) 
\end{equation}
where $\beta$ is an infinitesimal parameter to be set to $0$ at the end of the calculation.   

As discussed in Ref.~\cite{Amore16}, the basis set of eigenfunctions of $\hat{H}_0$ is 
\begin{eqnarray}
\Psi_{p,n}(x,y) = \psi_n(y)  \otimes \left\{
\begin{array}{ccc}
\phi_o(x) &,&  {\rm ground \ state} \, , \nonumber \\
\phi_p^{(e)}(x) &,& {\rm even} \, ,  \nonumber \\
\phi_p^{(o)}(x) &,& {\rm odd} \, , \nonumber
\end{array}
\right .
\end{eqnarray}
where 
\begin{eqnarray}
\phi_0(x) &=& \sqrt{\beta} e^{-\beta |x|} \, , \nonumber \\
\phi_p^{(e)}(x)  &=& \frac{\sqrt{2}}{\sqrt{p^2+ \beta^2}} \left[ p \cos (p x)-\beta \sin(p|x|) \right] \, , \nonumber \\
\phi_p^{(o)}(x)  &=& \sqrt{2}  \sin(p x) \, , \nonumber
\end{eqnarray}
and 
\begin{eqnarray}
\psi_n(y) &=& \sqrt{\frac{2}{b}} \sin \left[\frac{n \pi}{b} (y+b/2)\right] \, . \nonumber
\end{eqnarray}

The eigenvalues of $\hat{H}_0$ are~\footnote{Notice that $\epsilon_{0,1} = -\beta^2 +\frac{\pi^2}{b^2} < \frac{\pi^2}{b^2}$ and therefore
it is separated from the continuum. }
\begin{eqnarray}
\epsilon_{0,n} &=& -\beta^2 +\frac{n^2\pi^2}{b^2} \, , \nonumber  \\
\epsilon^{(e)}_{p,n} &=& \epsilon^{(o)}_{p,n} = p^2 +\frac{n^2\pi^2}{b^2} \, . \nonumber
\end{eqnarray}

We find convenient to introduce the Dirac notation $|0,n\rangle$, $|p^{(e)},n\rangle$ and $|p^{(o)},n\rangle$ to indicate the 
eigenstates of $\hat{H}_0$.

Using the explicit form of the eigenfunctions of $\hat{H}_0$ given above, one can work out the perturbative expressions for the
energy and, after taking the limit $\beta \rightarrow 0^+$, obtain the finite expressions given in Ref.~\cite{Amore16}:
\begin{eqnarray}
\lim_{\beta\rightarrow 0^+} E_0^{(1)} &=& 0 \\
\lim_{\beta\rightarrow 0^+} E_0^{(2)} &=&  - \frac{\pi^4}{b^6}  \left[ \int_{-\infty}^{\infty} dx \int_{-b/2}^{b/2} dy \ \sigma(x,y)  \cos ^2\left(\frac{\pi  y}{b}\right)    \right]^2 \\
\lim_{\beta \rightarrow 0^+} E_0^{(3)} &=&
\frac{2 \pi^6}{b^9} \left( \int_{-\infty}^{\infty} dx_3\int_{-b/2}^{b/2} dy_3
 \cos^2\left(\frac{\pi  y_3}{b}\right) \sigma \left(x_3,y_3\right)\right)
\nonumber \\
&\times&  \int_{-\infty}^{\infty} dx_1\int_{-b/2}^{b/2} dy_1\int_{-\infty}^{\infty} dx_2\int_{-b/2}^{b/2} dy_2 \Big[
\left| x_1-x_2\right| \sigma \left(x_1,y_1\right) \sigma \left(x_2,y_2\right) \nonumber \\
& &  \times \cos ^2\left(\frac{\pi  y_1}{b}\right)  \cos^2\left(\frac{\pi  y_2}{b}\right)
- b \cos \left(\frac{\pi  y_1}{b}\right) \cos \left(\frac{\pi  y_2}{b}\right)
\nonumber \\
& & \times \sigma \left(x_1,y_1\right) \sigma\left(x_2,y_2\right)  \left. \mathcal{G}_{2}^{(0)}(\mathbf{x}_1, \mathbf{x}_2) \right|_{\beta=0} \Big] \, .
\end{eqnarray}
where~\footnote{Notice that we have changed the notation of Ref.~\cite{Amore16} to allow referring to more general Green's functions.}
\begin{eqnarray}
\mathcal{G}_{0}^{(\ell)}(\mathbf{x}, \mathbf{x}') &\equiv&  \int_0^\infty \frac{dp}{2\pi} \frac{\phi_p(x) \phi_p(x') \psi_{1}(y) \psi_1(y')}{(\epsilon_{p,1} - \epsilon_{0,1})^{\ell+1}} \nonumber \\
\mathcal{G}_{1}^{(\ell)}(\mathbf{x}, \mathbf{x}') &\equiv&  {\sum_{n=2}^\infty}  \frac{\phi_0(x) \phi_0(x') \psi_{n}(y) \psi_n(y')}{(\epsilon_{0,n} - \epsilon_{0,1})^{\ell+1}} \nonumber \\
\mathcal{G}_{2}^{(\ell)}(\mathbf{x}, \mathbf{x}') &\equiv& {\sum_{n=2}^\infty} \int_0^\infty \frac{dp}{2\pi} \frac{\phi_p(x) \phi_p(x') \psi_{n}(y) \psi_n(y')}{(\epsilon_{p,n} - \epsilon_{0,1})^{\ell+1}}  \nonumber
\end{eqnarray}

Before discussing the fourth order, it is worth to comment that, as discussed in \cite{Amore16}, a bound state is present
only if the condition
\begin{eqnarray}
\int_{-\infty}^\infty \int_{-b/2}^{b/2} \sigma(x,y) \cos^2 \frac{\pi y}{b} dx dy >0 
\label{var_con}
\end{eqnarray}
is met.

We briefly review the discussion in Ref.~\cite{Amore16}: the condition (\ref{var_con})
can be derived calculating the Rayleigh quotient
\begin{eqnarray}
W = \frac{\langle \Psi | (- \Delta) | \Psi \rangle}{\langle \Psi | \Sigma  | \Psi \rangle}  \nonumber
\end{eqnarray}
using the variational function
\begin{eqnarray}
\Psi(x,y) = \sqrt{a} \ e^{-a |x|} \sqrt{\frac{2}{b}} \ \sin \frac{n \pi (y+b/2)}{b} \nonumber
\end{eqnarray}
and minimizing with respect to the variational parameter $a$:
\begin{eqnarray}
a_{min} \approx \frac{\pi^2}{b^3} \int_{-\infty}^{+\infty} \int_{-b/2}^{b/2} \sigma(x,y) \cos^2 \frac{\pi y}{b}  dxdy \nonumber
\end{eqnarray}
Given that, in order to obtain a bound state, $a$ must be positive, the condition (\ref{var_con}) follows.
   
In a similar way, one can derive the expression for the perturbative correction to the energy of the fundamental mode to fourth order;
we find
\begin{eqnarray}
E_0^{(4)} &=&  \langle\sigma \rangle^4 \epsilon_0-6 \langle\sigma \rangle^2 \langle\sigma \Omega \sigma \rangle \epsilon_0^2 \nonumber \\
&+& \left(2 \langle\sigma \Omega \sigma \rangle^2+4 \langle\sigma \rangle \langle\sigma \Omega \sigma \Omega \sigma \rangle-4 \langle\sigma \rangle^2 \langle\sigma \Omega \Omega \sigma \rangle\right) \epsilon_0^3 \nonumber \\
&+& \left(-\langle\sigma \Omega \sigma \Omega \sigma \Omega \sigma \rangle+\langle\sigma \Omega \sigma \rangle \langle\sigma \Omega \Omega \sigma \rangle+2 \langle\sigma \rangle \langle\sigma \Omega \Omega \sigma \Omega \sigma \rangle \right. \nonumber \\
&-& \left. \langle\sigma \rangle^2 \langle\sigma \Omega \Omega \Omega \sigma \rangle\right) \epsilon_0^4 
\end{eqnarray}

The perturbative expressions written above must be evaluated taking the limit $\beta \rightarrow 0^+$ at the end of the calculation. For this reason it is convenient to work on the expectation values which appear in the expression and expand them around $\beta =0$.

For example, in the simplest case we have
\begin{eqnarray}
\langle\sigma \rangle = \beta \int dxdy e^{-2\beta |x|} (\psi_1(y))^2 \sigma(x,y) =  
\sum_{n=1}^\infty \kappa^{(n)}_1 \beta^n \nonumber
\end{eqnarray}
The expressions for the remaining expectation values can be found in Appendix \ref{appb}. In particular, 
in Table \ref{tab1} the coefficients $\kappa_n^{(j)}$ are subdivided into two classes: those which only 
contain longitudinal contributions (left column) and those which contain both longitudinal 
and tranverse contributions (right column).

\begin{table}
\caption{Coefficients appearing in the expression of the energy of the fundamental mode up to fourth order in perturbation
theory. The coefficients on the right side contain contributions
also from the transversal modes. \label{tab1} }
\begin{center}
\begin{tabular}{|ccc|ccc|}
\hline
$\parallel$ &  &  &  $\parallel + \perp$ & & \\
\hline
$\kappa_1^{(1)}$  & $\kappa_1^{(2)}$  &                    &                   &                    &    \\
$\kappa_2^{(0)}$  &                   &                    &  $\kappa_2^{(1)}$ & $\kappa_2^{(2)}$   &                  \\
$\kappa_3^{(-2)}$ & $\kappa_3^{(-1)}$ & $\kappa_3^{(0)}$   &                   &                    &                  \\
$\kappa_4^{(-1)}$ &                    &                   &  $\kappa_4^{(0)}$ &                    & \\
$\kappa_5^{(-4)}$ & $\kappa_5^{(-3)}$ &  $\kappa_5^{(-2)}$ &                   &                    &     \\
$\kappa_6^{(-3)}$ & $\kappa_6^{(-2)}$ &  $\kappa_6^{(-1)}$ &                   &                    &     \\
$\kappa_7^{(-2)}$ &                  &                     & $\kappa_7^{(-1)}$ &  $\kappa_7^{(0)}$     &    \\
\hline
\end{tabular}
\end{center}
\end{table}

Upon substitution of these expressions in the perturbative contributions of the energy we have
\begin{eqnarray}
E_0^{(1)} &=& O(\beta)\\
E_0^{(2)} &=& -\epsilon_0^2  \kappa _2^{(0)} +O(\beta)\\
E_0^{(3)} &=& \epsilon_0^3 \frac{\kappa_1^{(1)} \kappa_3^{(-2)} -\kappa_4^{(-1)} }{\beta} +
\epsilon_0^3 \left[ \kappa_1^{(2)} \kappa_3^{(-2)}  +\kappa_1^{(1)} \kappa_3^{(-1)} -\kappa_4^{(0)} \right] + O(\beta)
\end{eqnarray}
and
\begin{eqnarray}
E_0^{(4)} &=& \eta_{4a} \left(\frac{\epsilon_0^4}{\beta ^2}-4 \epsilon_0^3\right)  + 
\eta_{4b} \frac{\epsilon_0^4 }{\beta } + \eta_{4c} \epsilon_0^3 + \eta_{4d} \epsilon_0^4  + O(\beta)
\end{eqnarray}
where
\begin{eqnarray}
\eta_{4a} &\equiv& \left(-\kappa _5^{(-4)} (\kappa_1^{(1)})^2+2 \kappa_6^{(-3)} \kappa_1^{(1)}+\kappa_2^{(0)} \kappa_3^{(-2)}-\kappa_7^{(-2)}\right) \nonumber \\
\eta_{4b} &\equiv& \left(-\kappa_5^{(-3)} (\kappa_1^{(1)})^2-2 \kappa_1^{(2)} \kappa_5^{(-4)} \kappa_1^{(1)}+2 \kappa_6^{(-2)} \kappa_1^{(1)}+
\kappa_2^{(1)} \kappa_3^{(-2)} \right. \nonumber \\
&+& \left. \kappa_2^{(0)} \kappa_3^{(-1)}+2 \kappa_1^{(2)} \kappa_6^{(-3)}-\kappa_7^{(-1)}\right) \nonumber \\
\eta_{4c} &\equiv& 2  \left((\kappa_2^{(0)})^2+2 \kappa_1^{(1)} \left(\kappa_4^{(-1)}-\kappa_1^{(1)} \kappa_3^{(-2)}\right)\right) \nonumber \\
\eta_{4d} &\equiv& \left(-\kappa_5^{(-2)} (\kappa_1^{(1)})^2-2 \kappa_1^{(3)} \kappa_5^{(-4)} \kappa_1^{(1)}
-2 \kappa_1^{(2)} \kappa_5^{(-3)} \kappa_1^{(1)}+2 \kappa_6^{(-1)} \kappa_1^{(1)} + \kappa_2^{(2)} \kappa_3^{(-2)} \right. \nonumber \\
&+& \left. \kappa_2^{(1)} \kappa_3^{(-1)} 
+  \kappa_2^{(0)} \kappa_3^{(0)}-(\kappa_1^{(2)})^2 \kappa_5^{(-4)}+2 \kappa_1^{(3)}
\kappa _6^{(-3)}+2 \kappa_1^{(2)} \kappa_6^{(-2)}-\kappa_7^{(0)}\right)\nonumber
\end{eqnarray}

Observe that the potentially divergent terms in $E_0^{(3)}$ and $E_0^{(4)}$ only depend on the contributions stemming from 
the longitudinal excitations. While it was already proved in Ref.~\cite{Amore16} that $E_0^{(3)}$ is finite for $\beta \rightarrow 0^+$,
as it can be checked explicitly using the results in \ref{appb}, it is straightforward to verify that 
$\eta_{4a} = \eta_{4b} = 0$. Therefore $E_0^{(4)}$ is finite for $\beta \rightarrow 0^+$, as expected.

Using the expressions in the Appendix we have
\begin{eqnarray}
\eta_{4c} &=& \frac{2}{b^4} \left( \int dx dy \ \cos^2\frac{\pi y}{b} \sigma(x,y)  \right)^4
\end{eqnarray}
and
\begin{eqnarray}
\eta_{4d} = \eta_{4d}^{\parallel} + \eta_{4d}^{\perp}
\end{eqnarray}
where $\eta_{4d}^{\parallel}$ contains only contributions from longitudinal modes while 
$\eta_{4d}^{\perp}$ contains contributions also from trasversal modes.

Their explicit expressions are~\footnote{The expression for $g_2^{(0,0)}(x_1,y_1,x_2,y_2)$ is reported in  \ref{appa}.}
\begin{eqnarray}
\eta_{4d}^{\parallel} &=& \frac{1}{b^4} \left(\int dx_1 dy_1 \int dx_2 dy_2  x_1 (2 x_2-x_1) \cos^2\left(\frac{\pi  y_1}{b}\right) \cos^2\left(\frac{\pi  y_2}{b}\right) \sigma(x_1,y_1) \sigma(x_2,y_2)  \right) \nonumber \\
&\times& \left( \int dx_3 dy_3  \cos^2\left(\frac{\pi  y_3}{b}\right) \sigma(x_3,y_3) \right)^2 \nonumber \\
&-& \frac{2}{b^4} \left(\int dx_1 dy_1 \int dx_2 dy_2 \int dx_3 dy_3  | x_1 -x_2| \cdot |x_2-x_3| 
\cos^2\left(\frac{\pi  y_1}{b}\right) \cos^2\left(\frac{\pi  y_2}{b}\right) \cos^2\left(\frac{\pi  y_3}{b}\right) \right. \nonumber \\
&\times& \left. \sigma(x_1,y_1) \sigma(x_2,y_2)  \sigma(x_3,y_3)  \right) \cdot \left( \int dx_4 dy_4  \cos^2\left(\frac{\pi  y_4}{b}\right) \sigma(x_4,y_4) \right) \nonumber \\ 
&-& \frac{1}{b^4} \left(\int dx_1 dy_1 \int dx_2 dy_2  | x_1 -x_2| 
\cos^2\left(\frac{\pi  y_1}{b}\right) \cos^2\left(\frac{\pi  y_2}{b}\right) \sigma(x_1,y_1) \sigma(x_2,y_2) \right)^2  \\
\eta_{4d}^{\perp} &=& \frac{1}{b^3}  \int dx_1dy_1 \int dx_2dy_2
\int dx_3dy_3 \int dx_4dy_4
\cos \left(\frac{\pi y_1}{b}\right)  \cos \left(\frac{\pi y_2}{b}\right) 
\cos^2\left(\frac{\pi y_3}{b}\right) \cos^2\left(\frac{\pi y_4}{b}\right) \nonumber \\
&\times& \sigma (x_1,y_1) \sigma (x_2,y_2) \sigma(x_3,y_3) \sigma(x_4,y_4) 
 (2 \left| x_1-x_3\right| + \left|x_3-x_4\right|  ) \ g_2^{(0,0)}(x_1,y_1,x_2,y_2) \nonumber \\
&-& \frac{2}{b^2} \left( \int dx_1dy_1 \int dx_2dy_2 \int dx_3dy_3 
\cos \left(\frac{\pi y_1}{b}\right) \cos \left(\frac{\pi y_3}{b}\right) 
g_2^{(0,0)}(x_1,y_1,x_2,y_2)   g_2^{(0,0)}(x_2,y_2,x_3,y_3) \right. \nonumber \\
&\times& \left. \sigma(x_1,y_1) \sigma(x_2,y_2) \sigma(x_3,y_3) \right) \times
 \left( \int dx_4 dy_4  \cos^2\left(\frac{\pi y_4}{b}\right) \sigma(x_4,y_4) \right) \nonumber \\
&-& \frac{1}{b^2} \left[\int dx_1dy_1 \int dx_2dy_2  
\cos \left(\frac{\pi y_1}{b}\right) \cos \left(\frac{\pi y_2}{b}\right) 
g_2^{(0,0)}(x_1,y_1,x_2,y_2)  
\sigma(x_1,y_1) \sigma(x_2,y_2) \right]^2
\end{eqnarray}

We may write the perturbative formulas obtained above in a more compact form as
\begin{eqnarray}
E_0^{(2)} &=& - \frac{\pi^2}{b^2} \Delta_2^2 \\
E_0^{(3)} &=& - 2 \frac{\pi^2}{b^2} \Delta_2 (\Lambda_1 - \Delta_3) \\
E_0^{(4)} &=&   -\frac{\pi^2}{b^2}  \left[ -2\Delta_2^4 - \Delta_2^2 \Delta_4 + 2 \Delta_2 \Delta_5 + \Delta_3^2 - 2 \Lambda_2 - \Delta_3 \Lambda_1
+ 2 \Delta_2 \Lambda_3 + \Lambda_1^2 \right]
\end{eqnarray}
where we have introduced the definitions
\begin{eqnarray}
\Delta_1 &\equiv&  \frac{\pi}{b^2} \int dxdy \sigma(x,y) \nonumber \\
\Delta_2 &\equiv&  \frac{\pi}{b^2} \int dxdy \sigma(x,y) \cos^2 \frac{\pi y}{b} \nonumber \\
\Delta_3 &\equiv& \frac{\pi^3}{b^5} \int dx_1dy_1 \int dx_2dy_2 \sigma(x_1,y_1) \sigma(x_2,y_2) |x_1-x_2| 
\cos^2 \frac{\pi y_1}{b} \cos^2 \frac{\pi y_2}{b} \nonumber \\
\Delta_4 &\equiv& \frac{\pi^4}{b^6} \int dx_1dy_1 \int dx_2dy_2 \sigma(x_1,y_1) \sigma(x_2,y_2) x_1 (2x_2-x1) 
\cos^2 \frac{\pi y_1}{b} \cos^2 \frac{\pi y_2}{b} \nonumber \\
\Delta_5 &\equiv& \frac{\pi^5}{b^8} \int dx_1dy_1 \int dx_2dy_2 \int dx_3dy_3 \sigma(x_1,y_1) \sigma(x_2,y_2)\sigma(x_3,y_3) |x_1-x_2|  |x_2-x_3| \nonumber \\
&\times& \cos^2 \frac{\pi y_1}{b} \cos^2 \frac{\pi y_2}{b} \cos^2 \frac{\pi y_3}{b}  \nonumber \\
\Lambda_1 &\equiv& \frac{\pi^3}{b^4} \int dx_1dy_1 \int dx_2dy_2 \sigma(x_1,y_1) \sigma(x_2,y_2) 
\cos \frac{\pi y_1}{b} \cos \frac{\pi y_2}{b} g_2^{(0,0)}(x_1,y_1,x_2,y_2)  \nonumber \\
\Lambda_2 &\equiv& \frac{\pi^6}{b^9} \int dx_1dy_1 \int dx_2dy_2 \int dx_3dy_3 \int dx_4dy_4 \sigma(x_1,y_1) \sigma(x_2,y_2) \sigma(x_3,y_3) \sigma(x_4,y_4) 
|x_1-x_3| \nonumber \\
&\times& \cos \frac{\pi y_1}{b} \cos \frac{\pi y_2}{b} \cos^2 \frac{\pi y_3}{b} \cos^2 \frac{\pi y_4}{b}  g_2^{(0,0)}(x_1,y_1,x_2,y_2)  \nonumber \\
\Lambda_3 &\equiv& \frac{\pi^5}{b^6} \int dx_1dy_1 \int dx_2dy_2 \int dx_3dy_3 \sigma(x_1,y_1) \sigma(x_2,y_2) \sigma(x_3,y_3) 
 \nonumber \\
&\times& \cos \frac{\pi y_1}{b}  \cos \frac{\pi y_3}{b}   g_2^{(0,0)}(x_1,y_1,x_2,y_2) g_2^{(0,0)}(x_2,y_2,x_3,y_3)   \nonumber 
\end{eqnarray}
where $\Delta_1$ is the total extra mass of the inhomogeneous waveguide.

The energy up to fourth order can then be arranged in the form
\begin{eqnarray}
\Delta E_0 &\approx& E_0^{(2)}  + E_0^{(3)} + E_0^{(4)} = - \frac{\pi^2}{b^2} \left\{  \left( \Delta_2 + (\Lambda_1- \Delta_3)^2 \right)^2 + 
\Gamma \right\}  
\end{eqnarray}
where 
\begin{eqnarray}
\Gamma \equiv \left[ - 2 \Delta_2^4 + \Delta_2 \Delta_3 - \Delta_2^2 \Delta_4 + 2 \Delta_2 \Delta5 - \Delta_3 \Lambda_1 -2 \Lambda_2 +2 \Delta_2 \Lambda_3\right]
\end{eqnarray}

When we apply the formulas above to the solvable model discussed in Ref.~\cite{Amore16} we obtain
\begin{eqnarray}
E_0^{(4)} &=& \frac{\sigma ^4 \left(90 \pi ^6 b^2 \delta ^4-23 \pi ^8 \delta^6\right)}{720 b^8} \nonumber 
\end{eqnarray}
which reproduces the exact expression for the fourth order contribution reported in Ref.~\cite{Amore16}.

\section{A solvable model}
\label{sec:model}

The case where the second and third order contributions vanish is particularly interesting
and it deserves a detailed discussion. This situation is analogous to the case
discussed by Exner and Vugalter in Ref.~\cite{Exner97} for a uniform, weakly deformed, waveguide.

As previously observed in Ref.~\cite{Amore16}  this occurs when the density obeys the property
\begin{eqnarray}
\int dx dy \ \cos^2\frac{\pi y}{b} \sigma(x,y)  = 0 \nonumber
\end{eqnarray}

In this limit the general formulas obtained in the previous section reduce to 
\begin{eqnarray}
\eta_{4c} &=& 0 \\
\eta_{4d}^{\parallel} &=&  - \frac{1}{b^4} \left(\int dx_1 dy_1 \int dx_2 dy_2  | x_1 -x_2| 
\cos^2\left(\frac{\pi  y_1}{b}\right) \cos^2\left(\frac{\pi  y_2}{b}\right) \right. \nonumber \\
&\times& \left. \sigma(x_1,y_1) \sigma(x_2,y_2) \right)^2  \\
\eta_{4d}^{\perp} &=& - \frac{1}{b^2} \left[\int dx_1dy_1 \int dx_2dy_2  
\cos \left(\frac{\pi y_1}{b}\right) \cos \left(\frac{\pi y_2}{b}\right) 
g_2^{(0,0)}(x_1,y_1,x_2,y_2)   \right. \nonumber \\
&\times& \left. \sigma(x_1,y_1) \sigma(x_2,y_2) \right]^2
\end{eqnarray}
and  the energy of the fundamental mode falls below the threshold of the continuum, 
signalling that the corresponding eigenfunction is localized in the region of the
heterogeneity.

\begin{figure}
\begin{center}
\bigskip\bigskip\bigskip
\includegraphics[width=6cm]{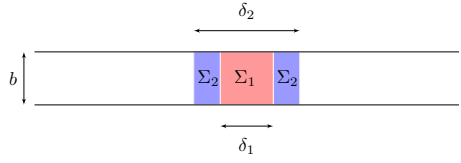}
\caption{(color online) Heterogeneous waveguide with three regions of different density.}
\label{Fig_1}
\end{center}
\end{figure}

To test this prediction, we consider a solvable model, represented by an infinite heterogeneous waveguide, parallel to  the horizontal axis and obeying Dirichlet boundary conditions on $y=\pm b/2$ (see Fig.~\ref{Fig_1}).

The density is  
\begin{eqnarray}
\Sigma(x) = \left\{  \begin{array}{ccc}
1+\sigma_1 & , & |x| < \delta_1/2 \\
1+\sigma_2 & , & \delta_1/2< |x| < \delta_2/2 \\
1 & , & |x| > \delta_2/2 \\
\end{array}
\right. \nonumber
\end{eqnarray}
where $\delta_2 \geq \delta_1 \geq 0$ (for $\sigma_1 = \sigma_2$ this problem 
reduces to the one discussed in Ref.~\cite{Amore16}).

We look for the solution to the Helmholtz equation
\begin{eqnarray}
-\Delta\Psi(x,y) = E \Sigma(x) \Psi(x,y) \nonumber
\end{eqnarray}
in the form
\begin{eqnarray}
\Psi(x,y)  = \sqrt{\frac{2}{b}} \sin \frac{\pi n (y+b/2)}{b} \times \left\{ \begin{array}{ccc}
A_1 \cos (p_1 x) & ,  &  |x|<\delta_1/2 \\
A_2 \cos (p_2 x+ q_2) & ,  &  \delta_1/2<|x|<\delta_2/2 \\
A_3  e^{-\alpha |x|} & ,& |x|>\delta_2/2 \\
\end{array}
\right. \nonumber 
\end{eqnarray}
where the unknown coefficients are to be obtained enforcing the continuity of the 
solution and its derivative at $x= \delta_1/2$ and $x= \delta_2/2$ (since the solution for 
the fundamental mode must be even, the matching at $x= -\delta_1/2$ and $x= -\delta_2/2$ is automatic). Since we are interested only in the fundamental mode we may set $n=1$.

By asking that $\Psi(x,y)$ be a solution to the Helmholtz equation on each region we obtain
\begin{eqnarray}
p_1 &=& \sqrt{k^2 (1+\sigma_1) - \pi^2/b^2 } \nonumber \\
p_2 &=& \sqrt{k^2 (1+\sigma_2) - \pi^2/b^2 } \nonumber \\
\alpha &=& \sqrt{\pi^2/b^2 -k^2}\nonumber 
\end{eqnarray}

From the matching of the solutions we obtain the transcendental equations
\begin{eqnarray}
A_1 \cos \left( \frac{\delta_1 p_1}{2} \right) &=& A_2 \cos \left( \frac{\delta_1 p_2}{2}  +q_2\right) \nonumber \\
A_1  p_1 \sin\left( \frac{\delta_1 p_1}{2} \right) &=& A_2 p_2 \sin \left( \frac{\delta_1 p_2}{2}  +q_2\right) \nonumber \\
A_3 e^{-\alpha \delta_2/2} &=& A_2 \cos \left( \frac{\delta_2 p_2}{2}  +q_2\right) \nonumber \\
- \alpha A_3 e^{-\alpha \delta_2/2} &=& A_2 p_2 \sin \left( \frac{\delta_2 p_2}{2}  +q_2\right) \nonumber 
\end{eqnarray}
which can be reduced to 
\begin{eqnarray}
\frac{p_1}{p_2} \tan \left( \frac{\delta_1 p_1}{2} \right) &=&  \tan \left( \frac{\delta_1 p_2}{2} +q_2 \right) \label{te1} \\
\alpha &=&  p_2  \tan \left( \frac{\delta_1 p_2}{2} +q_2 \right)  \label{te2}
\end{eqnarray}
after eliminating the amplitudes. 

We look for a solution to these equations, in the limit of weak inhomogeneities: to perform the appropriate
expansion in the density we introduce a parameter $\eta$, to keep track of the order of the expansion and
make the substitutions $\sigma_i \rightarrow \eta \sigma_i$ (at the end of the calculation
we will let $\eta \rightarrow 1$).

We also express $k$ and $q_2$ in terms of appropriate power series:
\begin{eqnarray}
q_2 = \sum_{n=0}^\infty c_n \eta^{n/2} \ \ \ ; \ \ \
k = \sqrt{\frac{\pi^2}{b^2} + \sum_{n=1}^\infty \kappa_n \eta^n} \nonumber
\end{eqnarray}

After substituting these expressions in the equations (\ref{te1}) and (\ref{te2}) one obtains the explicit
expression for the lowest eigenvalue
\begin{eqnarray}
E_0 &=& k^2 = \frac{\pi^2}{b^2} 
-\frac{\pi ^4 \left(\delta _1 \left(\sigma _1-\sigma_2\right)+\delta _2 \sigma _2\right)^2}{4 b^4} \nonumber \\
&+& \frac{\pi ^6 \left(\delta _1 \left(\sigma _1-\sigma
   _2\right)+\delta _2 \sigma _2\right) \left(\delta
   _1^3 \left(2 \sigma _1^2-3 \sigma _2 \sigma
   _1+\sigma _2^2\right)+3 \delta _2^2 \delta _1
   \left(\sigma _1-\sigma _2\right) \sigma _2+2
   \delta _2^3 \sigma _2^2\right)}{24 b^6} \nonumber \\
&+& \left[ \frac{\sigma _1^4 \left(90 \pi ^6 b^2 \delta _1^4-23
   \pi ^8 \delta _1^6\right)}{720 b^8}
   +\frac{\pi ^6 \delta _1^3 \left(\delta _1-\delta _2\right) \sigma _2 \sigma _1^3 \left(\pi ^2 \left(26
   \delta _1^2+15 \delta _2 \delta _1+5 \delta
   _2^2\right)-120 b^2\right)}{240 b^8} \right. \nonumber \\
&-& \left. \frac{\pi ^6
   \delta _1^2 \left(\delta _1-\delta _2\right)^2
   \sigma _2^2 \sigma _1^2 \left(\pi ^2 \left(79
   \delta _1^2+86 \delta _2 \delta _1+51 \delta
   _2^2\right)-432 b^2\right)}{576 b^8} \right. \nonumber \\
&+& \left. \frac{\pi ^6
   \delta _1 \left(\delta _1-\delta _2\right)^3
   \sigma _2^3 \sigma _1 \left(\pi ^2 \left(37
   \delta _1^2+56 \delta _2 \delta _1+47 \delta
   _2^2\right)-240 b^2\right)}{480 b^8}\right. \nonumber \\
&-& \left. \frac{\pi ^6
   \left(\pi ^2 \left(\delta _1-\delta _2\right)^4
   \left(47 \delta _1^2+86 \delta _2 \delta _1+92
   \delta _2^2\right) \sigma _2^4-360 b^2
   \left(\delta _2 \sigma _2-\delta _1 \sigma
   _2\right)^4\right)}{2880 b^8} \right] + \dots \nonumber
\end{eqnarray}
subject to the condition
\begin{eqnarray}
\delta _1 \left(\sigma _1-\sigma_2\right)+\delta _2 \sigma _2 \geq 0  \nonumber
\end{eqnarray}

In particular it is interesting to consider the case $\sigma _1 = \frac{\left(\delta _1-\delta _2\right) \sigma _2}{\delta _1}$,
corresponding to a waveguide where the heterogeneity averages to zero; in this case the energy reduces to 
\begin{eqnarray}
E_0= \frac{\pi ^2}{b^2}-\frac{\pi ^8 \left(\delta _1-\delta _2\right)^4 \delta _2^2 \sigma_2^4}{576 b^8} + 
\frac{\pi ^{10} \left(\delta _1-3 \delta_2\right) \left(\delta _1-\delta_2\right)^5 \delta _2^2 \sigma _2^5}{5760 b^{10}} + \dots
\label{exact}
\end{eqnarray}
where we have reported the fifth order as well (we do not report the fifth order for the general case, because of its length).

For this model the perturbative formulas derived  in the previous section up to fourth order 
yield
\begin{eqnarray}
E_0^{(pert)}  &=& \frac{\pi ^2}{b^2} - \frac{\pi^8}{b^{12}} \left(\int dx_1 dy_1 \int dx_2 dy_2  | x_1 -x_2| 
\cos^2\left(\frac{\pi  y_1}{b}\right) \cos^2\left(\frac{\pi  y_2}{b}\right) \right. \nonumber \\
&\times& \left. \sigma(x_1,y_1) \sigma(x_2,y_2) \right)^2 \nonumber \\
&=& \frac{\pi ^2}{b^2}-\frac{\pi ^8 \left(\delta _1-\delta _2\right){}^4 \delta _2^2 \sigma_2^4}{576 b^8}
\end{eqnarray}
which confirms the exact result of eq.~(\ref{exact}).

\begin{figure}
\begin{center}
\bigskip\bigskip\bigskip
\includegraphics[width=6cm]{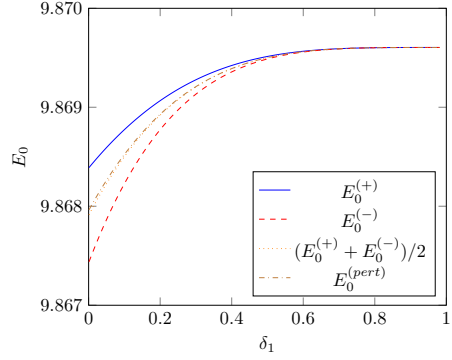}
\caption{(color online) Energy of the fundamental mode of the solvable model, for the case $\delta_2 = 1$, $b=1$, 
$\sigma _1 = \frac{\left(\delta _1-\delta _2\right) \sigma _2}{\delta _1}$ and $|\sigma_2|=1/10$.}
\label{Fig_2}
\end{center}
\end{figure}

In Fig.~\ref{Fig_2} we plot the energy of the fundamental mode for the case $\delta_2 = 1$, $b=1$, 
$\sigma _1 = \frac{\left(\delta _1-\delta _2\right) \sigma _2}{\delta _1}$, as a function of $\delta_1$.
$E_0^{(\pm)}$ correspond to the numerical solution of the equations (\ref{te1}) and (\ref{te2}) for $\sigma_2 = \pm 1/10$, 
while $E_0^{(pert)}$ is the expression  of Eq.~(\ref{exact}). Notice that, while $E_0^{(\pm)}$ departs from the 
perturbative formula $E_0^{(pert)}$ for $\delta_1 \rightarrow 0$, the average of the two is remarkably close to 
$E_0^{(pert)}$. This is consistent with the form of the fifth order contribution reported in Eq.~(\ref{exact}), which 
changes sign in the two cases.

\section{Conclusions}
\label{sec:Concl}

In this paper we have applied the method described in Ref.~\cite{Amore16} to calculate the fourth order perturbative
correction to the energy of the ground state of an infinite waveguide, with a small heterogeneity localized around a 
given internal point.

We may summarize the main results with the following points
\begin{itemize}
\item the expression for $E_0^{(4)}$ is finite for $\beta \rightarrow 0^+$, as expected (notice that, as the perturbative
order increases there are more potentially divergent terms; for instance, while the third order only contains a 
term which diverges as $1/\beta$, the fourth order contains a term that diverges as $1/\beta^2$ as well);
\item for waveguides where the second and third orders vanish, there may still be a bound state and the
energy gap scales as the fourth power in the density (consistent with the observation made in ref.~\cite{Exner97} for the problem
of the deformed waveguide);
\item the exact results for two solvable models are reproduced to fourth order; 
\item the perturbative scheme adopted in this paper and in Ref.~\cite{Amore16} is fully consistent, and it could 
be used to obtain  higher order contributions; 
\end{itemize}

\section*{Acknowledgements}
This research was supported by the Sistema Nacional de Investigadores (M\'exico).
The figures were produced using  Tikz \cite{tikz}.

\appendix

\section{Green's function}
\label{appa}

In this Appendix we derive the relevant properties of the Green's functions needed in the calculation, and work out the
leading behavior for $\beta \rightarrow 0^+$.

We define the operator
\begin{eqnarray}
\hat{\Omega}_\gamma \equiv \left[{\sum_{n=2}^\infty}  \frac{1}{\epsilon_{0,n} - \epsilon_{0,1} + \gamma}  | 0,n \rangle \langle 0,n |
+ {\sum_{n=1}^\infty} \int_0^\infty \frac{dp}{2\pi} \frac{1}{\epsilon_{p,n} - \epsilon_{0,1} + \gamma } | p,n \rangle \langle p,n | \right]
\nonumber
\end{eqnarray}
and expand it around $\gamma=0$ as
\begin{eqnarray}
\hat{\Omega}_\gamma = \sum_{\ell =0}^\infty (-1)^\ell \hat{\Omega}^{(\ell+1)} \gamma^\ell \nonumber
\end{eqnarray}
where
\begin{eqnarray}
\hat{\Omega}^{(\ell+1)}  \equiv \left[{\sum_{n=2}^\infty}  \frac{1}{(\epsilon_{0,n} - \epsilon_{0,1})^{\ell+1}}  | 0,n \rangle \langle 0,n |
+ {\sum_{n=1}^\infty} \int_0^\infty \frac{dp}{2\pi} \frac{1}{(\epsilon_{p,n} - \epsilon_{0,1})^{\ell+1}} | p,n \rangle \langle p,n | \right]
\nonumber
\end{eqnarray}

Notice that $\hat{\Omega}^{(\ell+1)}$  obey the relations
\begin{eqnarray}
(\hat{H}_0 - \epsilon_{0,1}) \hat{\Omega}^{(1)} &=& \hat{1} -  | 0,1 \rangle \langle 0,1 | \nonumber \\
(\hat{H}_0 - \epsilon_{0,1}) \hat{\Omega}^{(\ell+1)} &=& \hat{\Omega}^{(\ell)}\nonumber
\end{eqnarray}

We define the 
\begin{eqnarray}
G_\gamma (\mathbf{x}_1, \mathbf{x}_2) &\equiv& \langle \mathbf{x_1} | \hat{\Omega}_\gamma | \mathbf{x_2} \rangle \nonumber \\
&=& \left[{\sum_{n=2}^\infty}  \frac{\phi_0(x_1) \phi_0(x_2) \psi_{n}(y_1) \psi_n(y_2)}{\epsilon_{0,n}-\epsilon_{0,1} +\gamma }
+ {\sum_{n=1}^\infty} \int_0^\infty \frac{dp}{2\pi} \frac{\phi_p(x_1) \phi_p(x_2) \psi_{n}(y_1) \psi_n(y_2)}{\epsilon_{p,n}-\epsilon_{0,1} +\gamma } \right] \nonumber \\
&=& \int_0^\infty \frac{dp}{2\pi} \frac{\phi_p(x_1) \phi_p(x_2) \psi_{1}(y_1) \psi_1(y_2)}{\epsilon_{p,1}-\epsilon_{0,1}+\gamma}  \nonumber \\
&+&\left[{\sum_{n=2}^\infty}  \frac{\phi_0(x_1) \phi_0(x_2) \psi_{n}(y_1) \psi_n(y_2)}{\epsilon_{0,n}-\epsilon_{0,1}+\gamma }
+ {\sum_{n=2}^\infty} \int_0^\infty \frac{dp}{2\pi} \frac{\phi_p(x_1) \phi_p(x_2) \psi_{n}(y_1) \psi_n(y_2)}{\epsilon_{p,n}-\epsilon_{0,1}+\gamma} \right] \nonumber \\
&\equiv& G_{\gamma 0}(\mathbf{x}_1, \mathbf{x}_2) + G_{\gamma 1}(\mathbf{x}_1, \mathbf{x}_2)  + G_{\gamma 2}(\mathbf{x}_1, \mathbf{x}_2)
\nonumber
\end{eqnarray}

We have
\begin{eqnarray}
G_{\gamma i} (\mathbf{x}_1, \mathbf{x}_2) = \sum_{\ell = 0}^\infty (-1)^\ell \mathcal{G}_{i}^{(\ell)}(\mathbf{x}_1, \mathbf{x}_2) \ \gamma^\ell
\nonumber
\end{eqnarray}
with $i=0,1,2$.

Clearly the integrals in the first and third Green's functions can be performed using the residue theorem; for example, 
after evaluating $G_{\gamma 0} (\mathbf{x}, \mathbf{x}')$  in this way, and expanding in $\gamma$, one finds
\begin{eqnarray}
\mathcal{G}_{0}^{(0)}(\mathbf{x}_1, \mathbf{x}_2) &=& \cos \left(\frac{\pi {y_1}}{b}\right) \cos \left(\frac{\pi {y_2}}{b}\right) 
\left\{ \frac{1}{2 b \beta} 
-\frac{1}{2b}   \left(\left| x_1\right| +\left| x_2\right| +2 \left|x_1-x_2\right|  \right) + \dots \right\} \nonumber \\
\mathcal{G}_{0}^{(1)}(\mathbf{x}, \mathbf{x}') &=& \cos \left(\frac{\pi {y_1}}{b}\right) \cos \left(\frac{\pi {y_2}}{b}\right) 
\left\{ \frac{1}{8 \beta ^3}
-\frac{  (\left| x_1\right| +\left| x_2\right| )}{8 b \beta ^2} + 
\frac{2 \left| {x_1}\right|  \left| {x_2}\right| -3 {x_1}^2+8 {x_1} {x_2}-3 {x_2}^2}{16 b \beta } \right. \nonumber \\
&+& \left. \frac{3 \left| {x_1}\right|  \left({x_1}^2+3 {x_2}^2\right)+3 \left| {x_2}\right|  \left(3
   {x_1}^2+{x_2}^2\right)+8 |{x_1}-{x_2}|^3 }{48 b} + \dots \right\} \nonumber \\
\mathcal{G}_{0}^{(2)}(\mathbf{x}_1, \mathbf{x}_2) &=&  \frac{1}{2}\ \cos \left(\frac{\pi {y_1}}{b}\right) \cos \left(\frac{\pi {y_2}}{b}\right)
\left\{ \frac{1}{8 b \beta ^5} -\frac{\left| {x_1}\right| +\left| {x_2}\right| }{8 b \beta ^4}
-\frac{-2 \left| {x_1}\right|  \left| {x_2}\right| + {x_1}^2-4 {x_1} {x_2}+ {x_2}^2}{16 b \beta ^3} \right. \nonumber \\
&+& \left. \frac{\left| {x_1}\right|  \left({x_1}^2+3 {x_2}^2\right)+\left| {x_2}\right|  \left(3
   {x_1}^2+{x_2}^2\right)}{48 b \beta ^2} \right. \nonumber \\
&+& \left.  \frac{-4 \left| {x_1}\right|  \left| {x_2}\right|  \left({x_1}^2+{x_2}^2\right)+5 {x_1}^4-24 {x_1}^3
   {x_2}+30 {x_1}^2 {x_2}^2-24 {x_1} {x_2}^3+5 {x_2}^4}{192 b \beta }   \right. \nonumber \\
&-& \left. \frac{5 \left| {x_1}\right|  \left({x_1}^4+10 {x_1}^2 {x_2}^2+5 {x_2}^4\right)+5 \left| {x_2}\right| 
   \left(5 {x_1}^4+10 {x_1}^2 {x_2}^2+{x_2}^4\right)+16 \left| {x_1}-{x_2}\right|^5 }{960 b}   
+ \dots      \right\}   \nonumber
\end{eqnarray}

Notice that to obtain  $G_{\gamma 2} (\mathbf{x}, \mathbf{x}')$  one does not need to perform any calculation, since it can be obtained from
$G_{\gamma 0} (\mathbf{x}, \mathbf{x}')$ with the simple substitutions $\gamma \rightarrow \gamma + \frac{(n^2-1)\pi^2}{b^2}$ and $\psi_1(y) \rightarrow \psi_n(y)$ and summing over $n$. After expanding in $\gamma$ one has
\begin{eqnarray}
\mathcal{G}_{2}^{(0)}(\mathbf{x}_1, \mathbf{x}_2) &=& \sum_{j=0}^\infty g_2^{(0,j)} \beta^j 
=  \sum_{n=2}^\infty \sin \left(\frac{\pi  n \left(\frac{b}{2}+y_1\right)}{b}\right) \sin \left(\frac{\pi  n \left(\frac{b}{2}+y_2\right)}{b}\right) \frac{e^{-\frac{\pi  \sqrt{n^2-1} \left| x_1-x_2\right| }{b}}}{\pi  \sqrt{n^2-1}} + O(\beta ) \nonumber  \\
\mathcal{G}_{2}^{(1)}(\mathbf{x}_1, \mathbf{x}_2)  &=& \sum_{j=0}^\infty g_2^{(0,j)} \beta^j =  \sum_{n=2}^\infty \sin \left(\frac{\pi  n \left(\frac{b}{2}+y_1\right)}{b}\right) \sin \left(\frac{\pi  n \left(\frac{b}{2}+y_2\right)}{b}\right)  \frac{b e^{-\frac{\pi  \sqrt{n^2-1} \left| x_2-x_1\right| }{b}}}{2 \pi ^3 \left(n^2-1\right)^2} \nonumber \\
&\times& \left[\pi  \left(n^2-1\right) \left| x_1-x_2\right| +b \sqrt{n^2-1}\right] + O(\beta )  \nonumber \\
\mathcal{G}_{2}^{(2)}(\mathbf{x}_1, \mathbf{x}_2)  &=& \sum_{j=0}^\infty g_2^{(0,j)} \beta^j = 
\sum_{n=2}^\infty \sin \left(\frac{\pi  n \left(\frac{b}{2}+y_1\right)}{b}\right) \sin \left(\frac{\pi  n \left(\frac{b}{2}+y_2\right)}{b}\right)  
\frac{b^2 e^{-\frac{\pi  \sqrt{n^2-1} \left| x_1-x_2\right| }{b}}}{8 \pi ^5 \left(n^2-1\right)^3} \nonumber \\
&\times&  \left[ 3 \pi  b \left(n^2-1\right) \left|x_1-x_2\right| +3 b^2 \sqrt{n^2-1}+\pi ^2 \left(n^2-1\right)^{3/2} \left(x_1-x_2\right){}^2\right] + O(\beta ) \nonumber 
\end{eqnarray}

Finally it is easy to work out the leading $\beta$ dependence of $\mathcal{G}_{1}^{(\ell)}(\mathbf{x}, \mathbf{x}')$ for $\beta \rightarrow 0$:
\begin{eqnarray}
\mathcal{G}_{1}^{(\ell)}(\mathbf{x}_1, \mathbf{x}_2) &=&  {\sum_{n=2}^\infty}  \frac{\phi_0(x_1) \phi_0(x_2) \psi_{n}(y_1) \psi_n(y_2)}{(\epsilon_{0,n} - \epsilon_{0,1})^{\ell+1}} \nonumber \\ 
&=&  \left[\beta -\beta ^2 \left(\left| x_1\right| +\left| x_2\right| \right) + O(\beta^3 ) \right] \ 
 {\sum_{n=2}^\infty}  \frac{\psi_{n}(y_1) \psi_n(y_2)}{(\epsilon_{0,n} - \epsilon_{0,1})^{\ell+1}} \nonumber 
\end{eqnarray}

\section{Expectation values}
\label{appb}

Here we report the expressions for the expectation values appearing in the perturbative corrections to the energy, up 
to fourth order.

\begin{itemize}
\item $\langle\sigma \rangle$
\begin{eqnarray}
\langle\sigma \rangle = \beta \int dxdy e^{-2\beta |x|} (\psi_1(y))^2 \sigma(x,y) =  
\sum_{n=1}^\infty \kappa^{(n)}_1 \beta^n \nonumber
\end{eqnarray}
Therefore
\begin{eqnarray}
\kappa^{(1)}_1 &=& \frac{2}{b} \int dxdy \cos ^2\left(\frac{\pi  y}{b}\right) \sigma (x,y) \nonumber\\
\kappa^{(2)}_1 &=& -\frac{4}{b} \int dxdy \left| x\right|  \cos ^2\left(\frac{\pi  y}{b}\right) \sigma (x,y) \nonumber 
\end{eqnarray}

\item $\langle \sigma \Omega \sigma\rangle$

\begin{eqnarray}
\langle \sigma \Omega \sigma\rangle &\equiv&  \beta \int dx_1 dy_1 \int dx_2 dy_2 e^{-\beta (|x_1|+|x_2|)} \sigma(x_1,y_1)
\sigma(x_2,y_2) \psi_1(y_1) \psi_1(y_2) \nonumber \\
&\times& \left[ \mathcal{G}_0^{(0)}(x_1,y_1,x_2,y_2) + \mathcal{G}_1^{(0)}(x_1,y_1,x_2,y_2)+ \mathcal{G}_2^{(0)}(x_1,y_1,x_2,y_2) \right]
\nonumber \\
&=&  \sum_{n=0}^\infty \kappa^{(n)}_2 \beta^n \nonumber
\end{eqnarray}

\begin{eqnarray}
\kappa^{(0)}_2 &=&  \frac{1}{b^2}  \int dx_1 dy_1 \int dx_2 dy_2 \ \cos ^2\left(\frac{\pi  y_1}{b}\right) \cos^2\left(\frac{\pi  y_2}{b}\right) 
\sigma\left(x_1,y_1\right) \sigma\left(x_2,y_2\right) \nonumber\\
\kappa^{(1)}_2 &=& - \frac{2}{b^2} \int dx_1 dy_1 \int dx_2 dy_2 \   
 \cos ^2\left(\frac{\pi y_1}{b}\right) \cos ^2\left(\frac{\pi  y_2}{b}\right) \sigma\left(x_1,y_1\right) \sigma\left(x_2,y_2\right)  \nonumber \\
&\times& \left(\left| x_1\right| +\left| x_1-x_2\right| +\left|x_2\right| \right) \nonumber \\ 
&+& \frac{2}{b} \int dx_1 dy_1 \int dx_2 dy_2 \cos\left(\frac{\pi  y_1}{b}\right) \cos\left(\frac{\pi  y_2}{b}\right)  
\sigma\left(x_1,y_1\right) \sigma\left(x_2,y_2\right) g_2^{(0,0)}(x_1,y_1,x_2,y_2)
\nonumber \\
\kappa^{(2)}_2 &=& \frac{1}{2 b^2} \int dx_1 dy_1 \int dx_2 dy_2 \   
   \cos ^2\left(\frac{\pi  y_1}{b}\right) \cos ^2\left(\frac{\pi y_2}{b}\right) 
   \sigma \left(x_1,y_1\right) \sigma\left(x_2,y_2\right) \nonumber \\
&\times&  \left[ \frac{\left(x_1+x_2\right) \left| x_1+x_2\right|  \left(x_1 \left|
   x_2\right| +x_2 \left| x_1\right| \right)}{2 x_1x_2} \right. \nonumber \\
   &+& \left. \frac{\left| x_1-x_2\right|  \left(\left(9 x_1-x_2\right) x_2 \left| x_1\right| -x_1 \left(x_1-9 x_2\right) \left| x_2\right|\right)}{2 x_1 x_2} \right. \nonumber \\
&+& \left. 10 \left| x_1\right|  \left|
   x_2\right| +7 x_1^2-4 x_2 x_1+7 x_2^2  \right] \nonumber \\
&+& \frac{2}{b} \int dx_1 dy_1 \int dx_2 dy_2 \cos \left(\frac{\pi  y_1}{b}\right) 
\cos \left(\frac{\pi  y_2}{b}\right) \sigma\left(x_1,y_1\right) \sigma\left(x_2,y_2\right) \nonumber \\
&\times& \left(-\left(\left|x_1\right| +\left| x_2\right| \right)
   g_2^{(0,0)}\left(x_1,y_1,x_2,y_2\right)+g_1^{(0,1)}\left(x_1,y_1,x_2,y_2\right)+
   g_2^{(0,1)}\left(x_1,y_1,x_2,y_2\right)\right) \nonumber
\end{eqnarray}

\item $\langle \sigma \Omega^2 \sigma\rangle$

\begin{eqnarray}
\langle \sigma \Omega^2 \sigma\rangle  &\equiv& 
\int dx_1 dy_1 \int dx_2 dy_2 \ \phi_o(x_1) \psi_1(y_1)   \sigma(x_1,y_1) \sigma(x_2,y_2)
\phi_o(x_2) \psi_1(y_2) \nonumber \\
&\times& \left[ \mathcal{G}_0^{(1)}(x_1,y_1,x_2,y_2) + \mathcal{G}_1^{(1)}(x_1,y_1,x_2,y_2)+ \mathcal{G}_2^{(1)}(x_1,y_1,x_2,y_2) \right]
\nonumber \\
&=&   \sum_{n=-2}^\infty \kappa^{(n)}_3 \beta^n \nonumber 
\end{eqnarray}
where
\begin{eqnarray}
\kappa^{(-2)}_3 &=&  \frac{1}{4 b^2} \int dx_1 dy_1 \int dx_2 dy_2  \ 
\cos ^2\left(\frac{\pi  y_1}{b}\right) \cos ^2\left(\frac{\pi  y_2}{b}\right)
   \sigma \left(x_1,y_1\right) \sigma \left(x_2,y_2\right) \nonumber \\
\kappa^{(-1)}_3 &=&  -\frac{1}{2 b^2} \int dx_1 dy_1 \int dx_2 dy_2  \ 
\cos^2\left(\frac{\pi  y_1}{b}\right) \cos ^2\left(\frac{\pi  y_2}{b}\right) \sigma
   \left(x_1,y_1\right) \sigma \left(x_2,y_2\right) \nonumber \\
&\times& \left(\left| x_1\right| +\left| x_2\right| \right)  \nonumber \\
\kappa^{(0)}_3 &=&   \frac{1}{b^2} \ \int dx_1 dy_1 \int dx_2 dy_2  \
\cos ^2\left(\frac{\pi  y_1}{b}\right) \cos ^2\left(\frac{\pi 
   y_2}{b}\right) \sigma \left(x_1,y_1\right) \sigma \left(x_2,y_2\right) \nonumber \\
&\times& \left( x_1 x_2+ \left|x_1\right|  \left| x_2\right| \right) \nonumber
\end{eqnarray}

\item $\langle \sigma \Omega \sigma \Omega \sigma\rangle$

\begin{eqnarray}
\langle \sigma \Omega \sigma \Omega \sigma\rangle  &\equiv& 
\int dx_1 dy_1 \int dx_2 dy_2 \int dx_3 dy_3  \ \phi_o(x_1) \psi_1(y_1)  \phi_o(x_3) \psi_1(y_3) \nonumber\\
&\times& \sigma(x_1,y_1) \sigma(x_2,y_2) \sigma(x_3,y_3) \ 
\left[ \mathcal{G}_{0}^{(0)}(\mathbf{x}_1, \mathbf{x}_2) + \mathcal{G}_{1}^{(0)}(\mathbf{x}_1, \mathbf{x}_2) + \mathcal{G}_{2}^{(0)}(\mathbf{x}_1, \mathbf{x}_2) \right] \nonumber \\
&\times& \left[ \mathcal{G}_{0}^{(0)}(\mathbf{x}_2, \mathbf{x}_3) + \mathcal{G}_{1}^{(0)}(\mathbf{x}_2, \mathbf{x}_3) +
\mathcal{G}_{2}^{(0)}(\mathbf{x}_2, \mathbf{x}_3) \right] \nonumber \\
&=&  \sum_{n=-1}^\infty \kappa^{(n)}_4 \beta^n \nonumber
\end{eqnarray}
where
\begin{eqnarray}
\kappa^{(-1)}_4 &=&  \frac{1}{2 b^3}  \int dx_1 dy_1 \int dx_2 dy_2   \int dx_3 dy_3  \
\cos ^2\left(\frac{\pi  y_1}{b}\right) \cos ^2\left(\frac{\pi  y_2}{b}\right) 
\cos^2\left(\frac{\pi  y_3}{b}\right)  \nonumber \\
&\times&  \sigma \left(x_1,y_1\right) \sigma\left(x_2,y_2\right) \sigma \left(x_3,y_3\right) \nonumber \\
\kappa^{(0)}_4 &=&  -\frac{1}{b^3} \int dx_1 dy_1 \int dx_2 dy_2   \int dx_3 dy_3  \
\cos ^2\left(\frac{\pi  y_1}{b}\right) \cos ^2\left(\frac{\pi  y_2}{b}\right)
   \cos ^2\left(\frac{\pi  y_3}{b}\right)  \nonumber \\
&\times&  \sigma \left(x_1,y_1\right) \sigma
   \left(x_2,y_2\right) \sigma \left(x_3,y_3\right) \ 
\left(\left| x_1\right| +\left| x_1-x_2\right| +\left|x_2\right| +\left| x_2-x_3\right| +\left| x_3\right|\right) 
\nonumber \\
&+&
\frac{2}{b^2} \int dx_1 dy_1 \int dx_2 dy_2   \int dx_3 dy_3  
\cos\left(\frac{\pi  y_1}{b}\right) \cos\left(\frac{\pi  y_2}{b}\right) 
\cos^2\left(\frac{\pi  y_3}{b}\right) \nonumber \\
&\times& \sigma\left(x_1,y_1\right) \sigma\left(x_2,y_2\right) 
\sigma\left(x_3,y_3\right)   g_2^{(0,0)}\left(x_1,y_1,x_2,y_2\right)
\end{eqnarray}

\item $\langle \sigma \Omega^3 \sigma\rangle$

\begin{eqnarray}
\langle \sigma \Omega^3 \sigma\rangle  &\equiv& 
\int dx_1 dy_1 \int dx_2 dy_2 \ \phi_o(x_1) \psi_1(y_1)   \sigma(x_1,y_1) \sigma(x_2,y_2)
\phi_o(x_2) \psi_1(y_2) \nonumber \\
&\times& \left[ \mathcal{G}_{0}^{(2)}(\mathbf{x}_1, \mathbf{x}_2) + \mathcal{G}_{1}^{(2)}(\mathbf{x}_1, \mathbf{x}_2) +
\mathcal{G}_{2}^{(2)}(\mathbf{x}_1, \mathbf{x}_2) \right] \nonumber \\
&=&  \sum_{n=-4}^\infty \kappa^{(n)}_5 \beta^n \nonumber 
\end{eqnarray}
where
\begin{eqnarray}
\kappa^{(-4)}_5 &=&  \frac{1}{8 b^2} \int dx_1 dy_1 \int dx_2 dy_2  \ \cos ^2\left(\frac{\pi  y_1}{b}\right) \cos ^2\left(\frac{\pi  y_2}{b}\right)
   \sigma \left(x_1,y_1\right) \sigma \left(x_2,y_2\right) \nonumber \\
\kappa^{(-3)}_5 &=&  - \frac{1}{4 b^2} \int dx_1 dy_1 \int dx_2 dy_2  \ 
\cos^2\left(\frac{\pi  y_1}{b}\right) \cos ^2\left(\frac{\pi  y_2}{b}\right) \sigma\left(x_1,y_1\right) \sigma \left(x_2,y_2\right) \nonumber\\
&\times& \left(\left| x_1\right| +\left| x_2\right| \right)  \nonumber \\
\kappa^{(-2)}_5 &=&  \frac{1}{8 b^2}  \int dx_1 dy_1 \int dx_2 dy_2  \
   \cos ^2\left(\frac{\pi  y_1}{b}\right) \cos ^2\left(\frac{\pi  y_2}{b}\right) \sigma \left(x_1,y_1\right) \sigma \left(x_2,y_2\right) 
\nonumber \\   
&\times&   \left(4 \left| x_1\right|  \left| x_2\right| +\left(x_1+x_2\right){}^2\right) \nonumber \\
\kappa^{(-1)}_5 &=&  - \frac{1}{4 b^2} \int dx_1 dy_1 \int dx_2 dy_2 
 \cos ^2\left(\frac{\pi  y_1}{b}\right) \cos ^2\left(\frac{\pi  y_2}{b}\right)
   \sigma \left(x_1,y_1\right) \sigma \left(x_2,y_2\right) \nonumber \\
&\times&   \left(x_1+x_2\right)
   \left(\left| x_2\right|  x_1+\left| x_1\right|  x_2\right) \nonumber \\
\kappa^{(0)}_5 &=&     \frac{1}{4 b^2} \ \int dx_1 dy_1 \int dx_2 dy_2 
\cos ^2\left(\frac{\pi  y_1}{b}\right) \cos ^2\left(\frac{\pi  y_2}{b}\right)
   \sigma \left(x_1,y_1\right) \sigma \left(x_2,y_2\right) \nonumber \\
&\times&   x_1 x_2 \left(\left| x_1\right|  \left| x_2\right| +x_1 x_2\right) \nonumber
\end{eqnarray}

\item $\langle \sigma \Omega^2 \sigma \Omega \sigma\rangle$

\begin{eqnarray}
\langle \sigma \Omega^2 \sigma \Omega \sigma\rangle  &\equiv& 
\int dx_1 dy_1 \int dx_2 dy_2 \int dx_3 dy_3  \ \phi_o(x_1) \psi_1(y_1) \phi_o(x_3) \psi_1(y_3) \nonumber\\
&\times& \sigma(x_1,y_1) \sigma(x_2,y_2) \sigma(x_3,y_3) \ 
\left[ \mathcal{G}_{0}^{(1)}(\mathbf{x}_1, \mathbf{x}_2) + \mathcal{G}_{1}^{(1)}(\mathbf{x}_1, \mathbf{x}_2) 
+ \mathcal{G}_{2}^{(1)}(\mathbf{x}_1, \mathbf{x}_2) \right] \nonumber \\
&\times& \left[ \mathcal{G}_{0}^{(0)}(\mathbf{x}_2, \mathbf{x}_3) + \mathcal{G}_{1}^{(0)}(\mathbf{x}_2, \mathbf{x}_3) +
\mathcal{G}_{2}^{(0)}(\mathbf{x}_2, \mathbf{x}_3) \right] \nonumber \\
&=&  \sum_{n=-3}^\infty \kappa^{(n)}_6 \beta^n \nonumber 
\end{eqnarray}
where
\begin{eqnarray}
\kappa^{(-3)}_6 &=& \frac{1}{8 b^3}  \int dx_1 dy_1 \int dx_2 dy_2 \int dx_3 dy_3  \ 
\cos ^2\left(\frac{\pi  y_1}{b}\right) \cos ^2\left(\frac{\pi  y_2}{b}\right) \cos^2\left(\frac{\pi  y_3}{b}\right) \nonumber \\
&\times& \sigma \left(x_1,y_1\right) \sigma\left(x_2,y_2\right) \sigma \left(x_3,y_3\right) \nonumber \\
\kappa^{(-2)}_6 &=& - \frac{1}{4 b^3} \int dx_1 dy_1 \int dx_2 dy_2 \int dx_3 dy_3  \
\cos^2\left(\frac{\pi y_1}{b}\right) \cos^2\left(\frac{\pi  y_2}{b}\right) \cos^2\left(\frac{\pi y_3}{b}\right)  \nonumber \\
&\times&  \sigma \left(x_1,y_1\right) \sigma \left(x_2,y_2\right) \sigma\left(x_3,y_3\right) 
\left(\left| x_1\right| +\left| x_2\right| +\left|x_2-x_3\right| +\left| x_3\right| \right)   \nonumber \\
&+&\frac{1}{4 b^2} \int dx_1 dy_1 \int dx_2 dy_2 \int dx_3 dy_3  \
\cos^2\left(\frac{\pi  y_1}{b}\right) \cos\left(\frac{\pi  y_2}{b}\right) 
\cos\left(\frac{\pi  y_3}{b}\right) \nonumber \\
&\times& g_2^{(0,0)}\left(x_2,y_2,x_3,y_3\right) \sigma
   \left(x_1,y_1\right) \sigma \left(x_2,y_2\right) \sigma \left(x_3,y_3\right) \nonumber\\
\kappa^{(-1)}_6 &=& \frac{1}{16 b^3} \int dx_1 dy_1 \int dx_2 dy_2   \int dx_3 dy_3  \
 \cos^2\left(\frac{\pi  y_1}{b}\right) \cos ^2\left(\frac{\pi  y_2}{b}\right) 
\cos^2\left(\frac{\pi  y_3}{b}\right) \nonumber \\
&\times&  \sigma\left(x_1,y_1\right) \sigma \left(x_2,y_2\right) \sigma \left(x_3,y_3\right)
\left(\frac{\left(x_2+x_3\right) \left| x_2+x_3\right|  \left(x_2 \left|
   x_3\right| +x_3 \left| x_2\right| \right)}{2 x_2 x_3} \right. \nonumber \\
&+& \left. \left(10 \left| x_2\right|  \left| x_3\right| +8 \left|x_1\right|  
   \left(\left| x_2\right| +\left| x_3\right|\right)+3 x_2^2+8 x_1 x_2-4 x_3 x_2+7 x_3^2\right)
\right. \nonumber \\
&-& \left. \frac{\left|x_2-x_3\right|  \left(x_2 \left(x_2-9 x_3\right) \left| x_3\right|
   -16 x_2 x_3 \left| x_1\right| +x_3 \left(x_3-9 x_2\right) \left|
   x_2\right| \right)}{2 x_2 x_3} \right)\nonumber \\
&+& \frac{1}{4 b^2}  \int dx_1 dy_1 \int dx_2 dy_2   \int dx_3 dy_3 
 \cos^2\left(\frac{\pi  y_1}{b}\right) \cos\left(\frac{\pi  y_2}{b}\right) 
 \cos\left(\frac{\pi  y_3}{b}\right) \nonumber \\
&\times& \sigma \left(x_1,y_1\right) \sigma\left(x_2,y_2\right)  \sigma\left(x_3,y_3\right) \nonumber \\
&\times& \left(-(2 \left|x_1\right|+\left|x_2\right|+\left|x_3\right|)  g_2^{(0,0)}\left(x_2,y_2,x_3,y_3\right)
 +g_1^{(0,1)}\left(x_2,y_2,x_3,y_3\right) \right. \nonumber \\
 &+& \left. g_2^{(0,1)}\left(x_2,y_2,x_3,y_3\right)\right)  \nonumber 
\end{eqnarray}

\item $\langle \sigma \Omega \sigma \Omega \sigma \Omega \sigma\rangle$

\begin{eqnarray}
\langle \sigma \Omega \sigma \Omega \sigma \Omega \sigma\rangle   &\equiv& 
\int dx_1 dy_1 \int dx_2 dy_2 \int dx_3 dy_3 \int dx_4 dy_4  \nonumber \\
&\times& \phi_o(x_1) \psi_1(y_1)  \phi_o(x_4) \psi_1(y_4) \sigma(x_1,y_1) \sigma(x_2,y_2) \sigma(x_3,y_3) \sigma(x_4,y_4)  \nonumber \\
&\times& \left[ \mathcal{G}_{0}^{(0)}(\mathbf{x}_1, \mathbf{x}_2) + \mathcal{G}_{1}^{(0)}(\mathbf{x}_1, \mathbf{x}_2) + \mathcal{G}_{2}^{(0)}(\mathbf{x}_1, \mathbf{x}_2) \right] \nonumber \\
&\times& \left[ \mathcal{G}_{1}^{(0)}(\mathbf{x}_2, \mathbf{x}_3) + \mathcal{G}_{1}^{(0)}(\mathbf{x}_2, \mathbf{x}_3) +
\mathcal{G}_{2}^{(1)}(\mathbf{x}_2, \mathbf{x}_3) \right] \nonumber \\
&\times& \left[ \mathcal{G}_{1}^{(0)}(\mathbf{x}_3, \mathbf{x}_4) + \mathcal{G}_{1}^{(0)}(\mathbf{x}_3, \mathbf{x}_4) +
\mathcal{G}_{2}^{(1)}(\mathbf{x}_3, \mathbf{x}_4) \right] \nonumber \\
&=&  \sum_{n=-2}^\infty \kappa^{(n)}_7 \beta^n \nonumber 
\end{eqnarray}

\begin{eqnarray}
\kappa^{(-2)}_7 &=& \frac{1}{4 b^4} \left( \int dx_1 dy_1 \cos ^2\left(\frac{\pi  y_1}{b}\right) 
\sigma \left(x_1,y_1\right)  \right)^4 \nonumber \\
\kappa^{(-1)}_7 &=& -\frac{1}{2 b^4} \ 
\int dx_1 dy_1\int dx_2 dy_2 \int dx_3 dy_3\int dx_4 dy_4 \nonumber \\
&\times& \cos ^2\left(\frac{\pi  y_2}{b}\right)  \cos ^2\left(\frac{\pi  y_3}{b}\right)
\cos^2\left(\frac{\pi  y_4}{b}\right) \cos ^2\left(\frac{\pi  y_1}{b}\right)  \nonumber \\
&\times& \sigma \left(x_1,y_1\right) \sigma \left(x_2,y_2\right) \sigma \left(x_3,y_3\right) \sigma\left(x_4,y_4\right) \nonumber \\
&\times& \left(\left| x_1\right| +\left| x_1-x_2\right| +\left| x_2\right| +\left|
   x_2-x_3\right| +\left| x_3\right| +\left| x_3-x_4\right| +\left| x_4\right| \right) \nonumber \\
&+& \frac{3}{2 b^3} \int dx_1 dy_1\int dx_2 dy_2 \int dx_3 dy_3\int dx_4 dy_4 \nonumber \\
&\times& \cos \left(\frac{\pi  y_1}{b}\right) \cos \left(\frac{\pi  y_2}{b}\right) 
 \cos^2\left(\frac{\pi  y_3}{b}\right) \cos^2\left(\frac{\pi  y_4}{b}\right)\nonumber \\
&\times&   g_2^{(0,0)}\left(x_1,y_1,x_2,y_2\right) \sigma \left(x_1,y_1\right) \sigma
   \left(x_2,y_2\right) \sigma \left(x_3,y_3\right) \sigma \left(x_4,y_4\right)
\end{eqnarray}

We omit writing the explicit expression for $\kappa^{(0)}_7$ because it is particularly lengthy.

\end{itemize}

\section*{Acknowledgements}
This research was supported by the Sistema Nacional de Investigadores (M\'exico).
The figures were produced using  Tikz \cite{tikz}.


\end{document}